%% file: artical003_v4_2023.11.20.tex
\documentclass[review]{elsarticle}

\usepackage{graphicx}
\usepackage{longtable}
\usepackage{float}

\usepackage{color}
\usepackage{soul}
\usepackage{bm}
\usepackage{makecell}
\usepackage{siunitx}
\usepackage{amsmath}
\usepackage{mathrsfs}
\usepackage{threeparttable}
\usepackage{multirow}
\usepackage{booktabs}
\sethlcolor{yellow}
\definecolor{r}{rgb}{1,0,0}

\usepackage[colorlinks,
linkcolor=red,
anchorcolor=blue,
citecolor=green
]{hyperref}

\begin{document}

\begin{frontmatter}

\title{Highly efficient and transferable interatomic potentials for $\alpha$-iron and $\alpha$-iron/hydrogen binary systems using deep neural networks}

\author[address1]{Shihao Zhang}
\author[address1]{Fanshun Meng}
\author[address2]{Rong Fu}
\author[address1]{Shigenobu Ogata\corref{correspondingauthor}}

\cortext[correspondingauthor]{Corresponding author}
\ead{ogata@me.es.osaka-u.ac.jp}

\address[address1]{Department of Mechanical Science and Bioengineering, Graduate School of Engineering Science, Osaka University, Osaka, 560-8531, Japan}
\address[address2]{School of Mechanical and Electronic Engineering, Lanzhou University of Technology, Lanzhou, People’s Republic of China}

\begin{abstract}
Artificial neural network potentials (NNPs) have emerged as effective tools for understanding atomic interactions at the atomic scale in various phenomena. Recently, we developed highly transferable NNPs for $\alpha$-iron and $\alpha$-iron/hydrogen binary systems (Physical Review Materials 5 (11), 113606, 2021). These potentials allowed us to investigate deformation and fracture in $\alpha$-iron under the influence of hydrogen.
However, the computational cost of the NNP remains relatively high compared to empirical potentials, limiting their applicability in addressing practical issues related to hydrogen embrittlement. In this work, building upon our prior research on iron-hydrogen NNP, we developed a new NNP that not only maintains the excellent transferability but also significantly improves computational efficiency (more than 40 times faster).
We applied this new NNP to study the impact of hydrogen on the cracking of iron and the deformation of polycrystalline iron. We employed large-scale through-thickness \{110\}$\langle 110 \rangle$ crack models and large-scale polycrystalline $\alpha$-iron models. The results clearly show that hydrogen atoms segregated at crack tips promote brittle-cleavage failure followed by crack growth. Additionally, hydrogen atoms at grain boundaries facilitate the nucleation of intergranular nanovoids and subsequent intergranular fracture.
We anticipate that this high-efficiency NNP will serve as a valuable tool for gaining atomic-scale insights into hydrogen embrittlement.
\end{abstract}


\end{frontmatter}


The degradation of mechanical performance in iron and steel due to hydrogen, two of the most commonly used structural materials, poses a significant challenge in various technological and industrial applications \cite{Hou2019,Gong2020}. To design novel iron-based alloys with superior resistance to hydrogen embrittlement, there is a growing demand for an atomic-scale understanding of hydrogen's interaction with defects within the iron matrix.
Experimental demonstrations of these atomic-scale mechanisms are challenging due to the high diffusivity, low solubility, and small size of hydrogen atoms \cite{Koyama2020}. Despite considerable efforts and successes, such as quantitative tests revealing hydrogen-enhanced dislocation motion in $\alpha$-iron \cite{Huang2023}, experimental validation remains limited. In this context, theoretical calculations and simulations, such as density functional theory (DFT) calculations and molecular dynamics (MD) simulations based on empirical interatomic potentials, offer valuable atom-level insights.
However, each of these methods has its inherent limitations. DFT calculations, while accurate, are computationally expensive. On the other hand, MD simulations using empirical interatomic potentials can suffer from questionable accuracy. For example, the empirical embedded atom method (EAM) potential often fails to accurately describe the core energy and structure of the 1/2$\langle 111 \rangle$ screw dislocation, which plays a dominant role in the plasticity of $\alpha$-iron \cite{Itakura2012,Ramasubramaniam2009}.

Recently, artificial neural network potential (NNP) provides an accurate tool to describe the atomic interactions \cite{Behler2021}, which yields both comparable accuracy and low computational cost comparing with DFT calculation. Within the framework of n2p2 high-dimensional NNP package \cite{Singraber2019a,Singraber2019}, we developed a general-purpose neural network interatomic potential for the $\alpha$-iron and hydrogen (Fe-H) binary system \cite{Meng2021}, which describes well the interaction of hydrogen with point defect, dislocation, surface, and tilt grain boundary (GB) in $\alpha$-iron, and can be applied to study at atomic level the hydrogen charging and discharging, hydrogen transportation, hydrogen trapping and desorption at defects, and so on. However, the NNP still suffers the high computational cost comparing with the empirical EAM potential. It much limits its application for many problems of practical interests in Fe-H system, such as, hydrogen-assisted cracking \cite{Song2013,Wan2019}, irradiation-induced dislocation loop \cite{Wan2014,Arakawa2011}, etc., for which a system size of ten thousands to millions of atoms, or even larger, is often required. To the best of our knowledge, to now, no attempt for Fe-H NNP has been performed to achieve comparable empirical potential efficiency together with DFT accuracy.

\begin{figure}[H]
	\centering
	\includegraphics[width=0.7\textwidth]{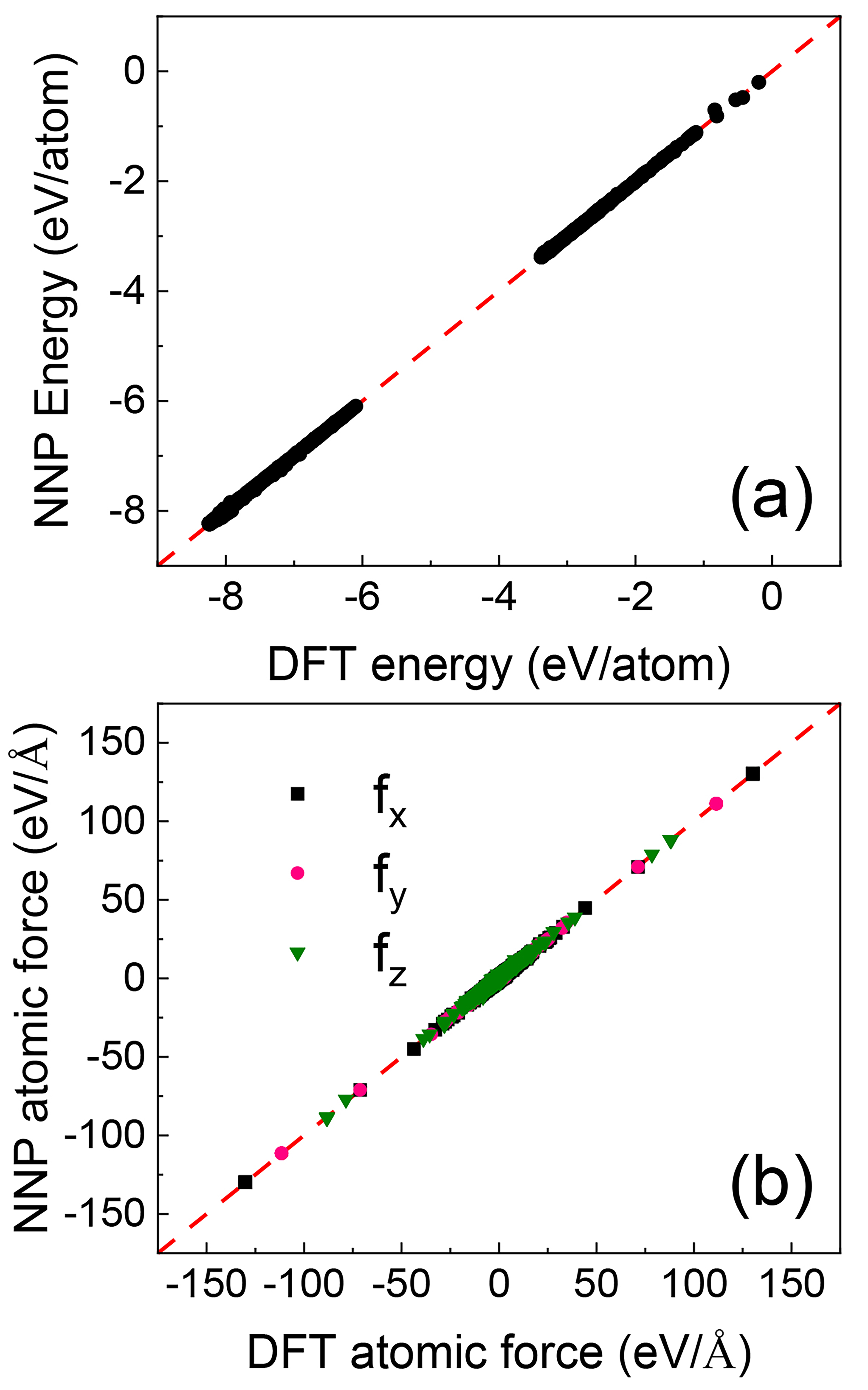}
	\caption{Comparison of DFT and NNP predicted (a) energies and (b) components of atomic forces along $x$-, $y$- and $z$-axes of the structures in the training dataset.}
	\label{Fig1}
\end{figure}

In this work, following our previous work of Fe-H NNP \cite{Meng2021}, a new NNP was developed and validated. With combination with the GPU accelerators, the new NNP shows much higher efficiency (more than 40 times faster) than the previous NNP with keeping the similar DFT accuracy. The deepMD-kit software package v2.1.5 \cite{Wang2018} was employed to construct the Fe-H NNP. A hybrid descriptor of se\_e2\_a and se\_e3 types was employed with the following configuration: cutoff radius $r_{\rm{cut}}=$ 6.5 \AA, configuration of embedding net \{30, 60, 120\}, and number of axis neurons 32 for se\_e2\_a type descriptor, as well as the cutoff radius $r_{\rm{cut}}=$ 5.0 \AA, and configuration of embedding net \{10, 20\} for se\_e3 type descriptor. The following fitting net of deepMD model in the training stage was used: the configuration of fitting net \{320, 320, 320\}, initial learning rate 10$^{-3}$, final learning rage 2.2$\times$10$^{-8}$, and training steps 3$\times$10$^6$. The prefactors of energy and force weights in loss functions are $p^{\rm start}_e = 0.02$, $p^{\rm limit}_e = 10$, $p^{\rm start}_f = 1000$, and $p^{\rm limit}_f = 1$, respectively. No virial data is available in the dataset, so the virial prefactors are set to 0, i.e., $p^{\rm start}_v = p^{\rm limit}_v = 0$, respectively. Same training dataset as our previous work \cite{Meng2021} was employed. Of the total 21928 DFT calculations, 20\% were randomly selected as a validation set which were used to determine whether over-fitting has occurred. The root mean square error of our Fe-H NNP for energy and atomic force are 4.8 meV/atom and 72.0 meV/{\AA} (see Fig. \ref{Fig1}), and are comparable with those of 3.0 meV/atom and 69.0 meV/{\AA} for the previous NNP \cite{Meng2021}, respectively. The performance of the new Fe-H NNP has been validated carefully, such as phonon dispersion, stacking fault energy, vacancy cluster formation energy, self-interstitial energy, trapping energy of H atom in vacancy, migration barrier of hydrogen-vacancy complex, adsorption energy for hydrogen atoms on low index surfaces, hydrogen-GB interaction, hydrogen-screw dislocation interaction, slip barrier of screw dislocation, kink nucleation energy of dislocation, hydrogen diffusivity in iron, and so on. See Table \ref{table1} and the Supplemental Material for the details of all the validation results.

\begin{figure}[H]
	\centering
	\includegraphics[width=0.9\textwidth]{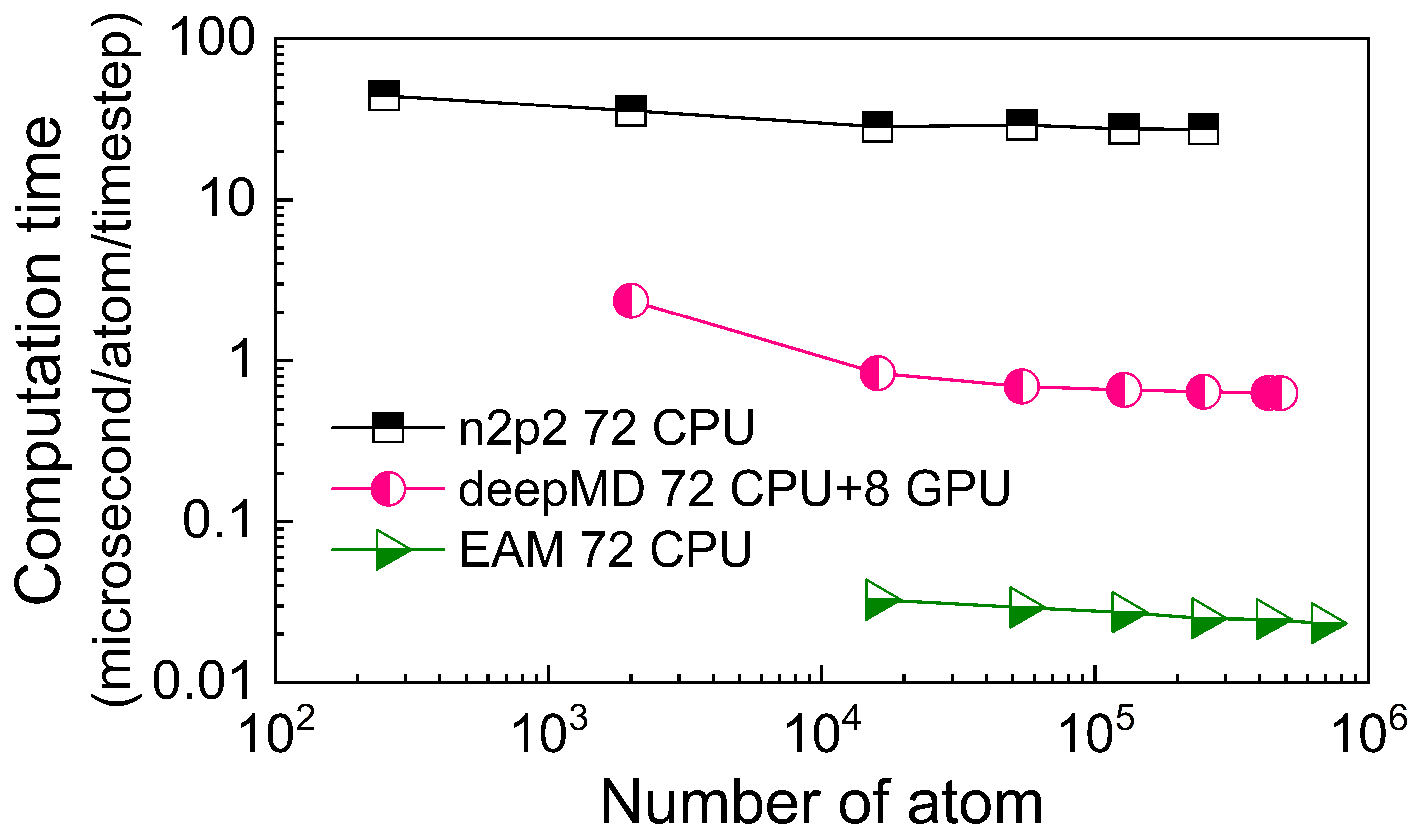}
	\caption{DeepMD NNP computational time per atom as function of the number of atoms running on single node of SQUID supercomputer at the Cybermedia Center of Osaka University \cite{squid} including two Intel Xeon Platinum 8368 and eight NVIDIA A100 GPU cores with LAMMPS code, comparing with EAM potentials proposed by Kumar et al. \cite{Kumar2023} and Wen et al. \cite{Wen2021} as well as our n2p2 NNP \cite{Meng2021}. To be noted that the n2p2 NNP currently doesn't support the GPU accelerator.}
	\label{Fig2}
\end{figure}

Fig. \ref{Fig2} illustrates the deepMD NNP computational time per atom as function of the number of atoms running on single node of SQUID supercomputer at the Cybermedia Center of Osaka University \cite{squid} including two Intel Xeon Platinum 8368 and eight NVIDIA A100 GPU cores with Large-scale Atomic/Molecular Massively Parallel Simulator (LAMMPS) code \cite{Plimpton1995}. The results of Fe-H EAM potentials proposed by Kumar et al. \cite{Kumar2023} and Wen et al. \cite{Wen2021} as well as our n2p2 NNP \cite{Meng2021} are also shown for comparison. It is seen that the deepMD NNP is more than 40 times faster that the n2p2 NNP with the help of GPU accelerator, e.g., 0.64 microsecond/atom/timestep of deepMD NNP vs 27.4 microsecond/atom/timestep of n2p2 NNP with hundred thousands atoms. Comparing with the empirical EAM potentials, the deepMD NNP is slower, but shows much better accuracy and transferability close to DFT calculation as demonstrated previously. Furthermore, the deepMD NNP could afford the MD simulation of up to 100 million atoms as presented in Ref. \cite{Lu2021} with multi nodes to fit in GPU memory.

\begin{figure}[H]
	\centering
	\includegraphics[width=1.0\textwidth]{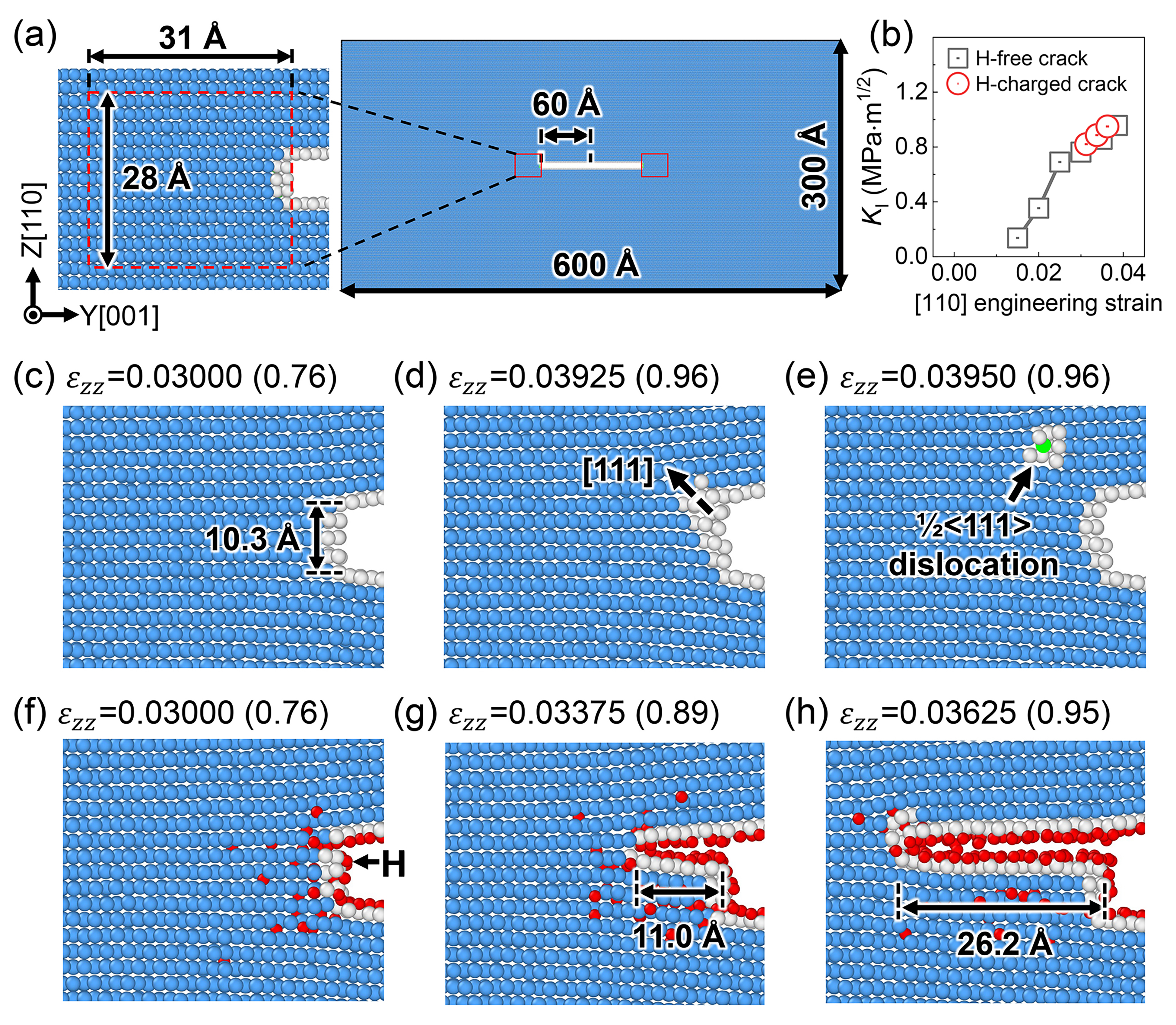}
	\caption{(a) Model of through-thickness \{110\}$\langle 110 \rangle$ crack and magnified view of the crack tip. The H-charged region are highlighted by red rectangle. The length of this quasi-two-dimensional model along the third direction (i.e., X-axis) is 8 {\AA}. (b) Relationship of $K_{\rm{I}}$ and tensile engineering strain $\varepsilon_{zz}$. See the Supplemental Material for the details of the determination of $K_{\rm{I}}$. To be noted that the difference of the H-free and H-charged models for $K_{\rm{I}}$ is that there is the new crack-tip for H-charged model owning to crack growth. Snapshot of the (c-e) H-free and (f-h) H-charged cracks under different tensile engineering strains $\varepsilon_{zz}$. All snapshots are displayed with the same regions as the zoom-in area in (a). The corresponding $K_{\rm{I}}$ are shown in the brackets (in MPa$\cdot$m$^{1/2}$). Fe atoms are colored according to the common neighbor analysis (CNA), i.e., blue is for perfect BCC lattice and gray for undetermined environment. H atom trapped at crack tip are shown by red.}
	\label{Fig4}
\end{figure}

To further demonstrate the performance of deepMD Fe-H NNP, the effect of H atoms on the crack propagation of the through-thickness \{110\}$\langle 110 \rangle$ crack was studied with a perfect $\alpha$-iron crystal model. The size is $X$: 8 {\AA} $\times$ $Y$: 600 {\AA} $\times$ $Z$: 300 {\AA}, consisting of approximately 126,000 atoms (see Fig. \ref{Fig4}a). The through-thickness crack with a half-length of 60 {\AA} was inserted by removing three layer atoms. The models are periodic in all directions. The LAMMPS code \cite{Plimpton1995} was employed for all the MD simulation. For the H-free crack, the mode-I loading normal to the crack plane at 300 K was performed by linearly rescaling the system length in the $z$-direction with a constant strain rate of $5 \times 10^8$ s$^{-1}$ in an NPT ensemble. The lateral stresses of the model were relaxed to zero. For the H-charged crack, H atoms were introduced via the hybrid MD/grand canonical Monte Carlo (GCMC) simulation with a constant H chemical potential of -2.2 eV corresponding to a bulk H concentration of 11.1 atomic ppm at 300 K. Constant time steps of 0.5 fs were used. 100 GCMC trials were conducted every 20 MD steps. The H-charged crack is initially deformed at $\varepsilon_{zz}$ = 0.03 ($K_{\rm I}$ = 0.76 MPa$\cdot$m$^{1/2}$, see Fig. \ref{Fig4}b) to provide a driving force for H segregation to the crack tip, and then same mode-I deformation was employed as the H-free crack. Hydrogen atoms were inserted using the hybrid MD/GCMC method specifically in the region near the crack tip as shown in Fig. \ref{Fig4}a. This approach was chosen because only the hydrogen atoms in proximity to the crack tip exhibit significant effects, and it accelerates the equilibrium of hydrogen concentration during mode-I loading. Atomic configurations were analyzed and visualized with the common neighbor analysis (CNA) \cite{Faken1994} and dislocation extraction algorithm (DXA) \cite{Stukowski2010} of the OVITO program \cite{Stukowski2009}. As shown in Figs. \ref{Fig4}c-e, for the H-free crack, the 1/2$\langle 111 \rangle$ dislocations are emitted from the crack tip, and the crack hardly propagated. It agrees with the results calculated by aenet NNP of pure iron \cite{Suzudo2022} and EAM potential \cite{Ersland2012}, as well as the experimental fact that BCC transition metals such as $\alpha$-Fe do not cleave along \{110\} plane \cite{Pineau2016}. In the case of H-charged crack, the H atoms prefer to be trapped at crack tip showing a strain/stress-driven segregation of hydrogen. As demonstrated in Fig. S1 in the Supplemental Material, the equilibrium bulk concentration of H increases exponentially with the [110] engineering strain under same chemical potentials, indicating that higher tensile strain/stress field attracts more H. Under mode-I loading, the localized strain/stress at crack tip increase with the increasing stress intensity factor ${K_{\rm{I}}}$, and more and more H atoms are trapped at crack tip. The aggregation of H permits brittle-cleavage failure followed by crack growth, and results in a ductile-to-brittle transition (see Figs. \ref{Fig4}f-h).

\begin{figure}[H]
	\centering
	\includegraphics[width=1.0\textwidth]{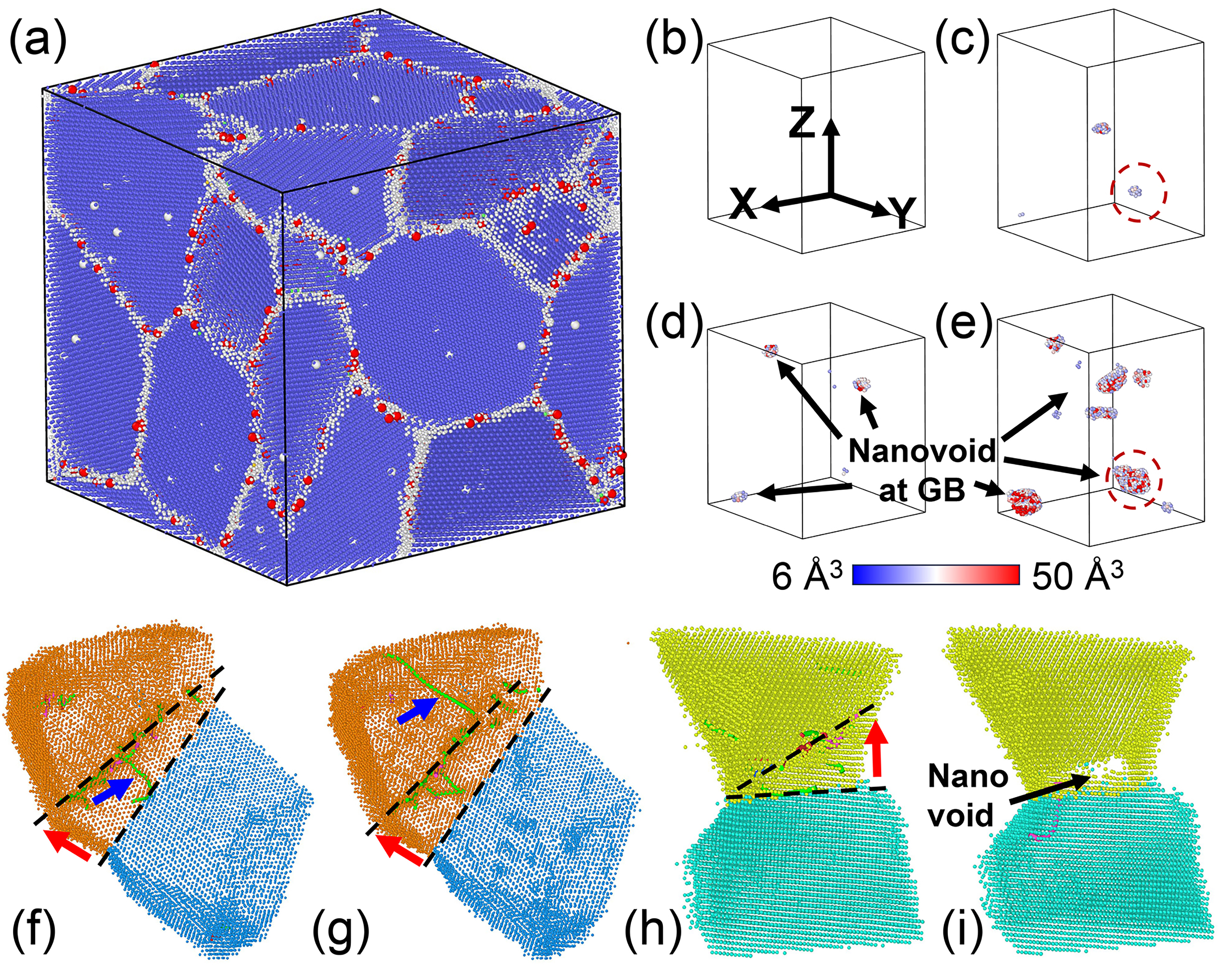}
	\caption{(a) Distribution of H atoms in polycrystalline model after heat treatment procedure. Fe atoms are colored according to the common neighbor analysis (CNA), i.e., blue is for perfect BCC lattice and gray for undetermined environment. The H atoms with Voronoi atomic volume larger than 7 \AA$^3$ are highlighted by red color, which clearly demonstrates the H segregation on GBs. (b-e) The snapshot of polycrystalline models: (b) H-free model, $\varepsilon_{zz} = $ 0.12, (c) H-free model, $\varepsilon_{zz} = $ 0.24, (d) H-charged model, $\varepsilon_{zz} = $ 0.12, and (e) H-charged model, $\varepsilon_{zz} = $ 0.24, where the Fe atoms are colored according to the Voronoi atomic volume. The atoms with a Voronoi volume less than 20 \AA$^3$ are not shown. (f-g) The snapshot of two grains in the (f) H-free model and (g) H-charged model with the tensile engineering strain $\varepsilon_{zz}$ = 0.087. (h-i) The snapshot of two grains where the nanovoid is nucleated as highlighted by red circle in (c,e) with the tensile engineering strain $\varepsilon_{zz} = $ 0.18 in the (h) H-free model and (i) H-charged model. Fe atoms within different grains are highlighted by different color. The H atoms are not shown. The observed nucleation of dislocation at GBs and the migration of GBs are highlighted by blue and red arrows, respectively. All the structures are shown with same orientation as (b).}
	\label{Fig3}
\end{figure}

In addition, a three-dimensional polycrystalline iron model with dimensions of 175 {\AA} $\times$ 175 {\AA} $\times$ 175 {\AA} was prepared (see Fig. \ref{Fig3}a). This model contains 10 randomly oriented grains and approximately 470,000 atoms. The structure of highly distorted GB in polycrystalline iron is believed to be well described by this NNP, for which many liquid configurations of pure iron have been included in the training dataset \cite{Meng2021}. A total of 2350 H atoms were randomly inserted in the polycrystalline model, corresponding to the H concentration of 5000 atomic ppm. A much high H concentration than the typical experimental value ($<$100 atomic ppm \cite{DaSilva1976}) was used with considering the increase of H concentration under tensile stress as shown in Fig. S1 in the Supplemental Material. Then, in NPT ensemble, the H-charged models were initially heated to 800 K for 100 ps, then cooled down to 300 K over the course of another 100 ps, followed by an additional annealing at 300 K for 100 ps. The purpose of this heat treatment procedure is to bring the structure of the GB and the distribution of hydrogen atoms close to equilibrium state. For the H-free model, the same heat treatment procedure was performed. As shown in Fig. \ref{Fig3}a, the H atoms segregate to the GB because of the strong H trapping effect of the GB at room temperature and the high diffusivity of H atoms in the $\alpha$-Fe lattice. The MD tensile simulation was then performed at 300 K with a constant strain rate of 6$\times$10$^8$ s$^{-1}$. A uniaxial tensile load was applied with the lateral stresses of the model relaxed to zero. As depicted in Figs. \ref{Fig3}f-g, we observe dislocation nucleation from GBs and GB migration as the plastic response to tensile deformation. Notably, the presence of hydrogen atoms segregated at GBs impedes GB migration, as illustrated in Figs. \ref{Fig3}h-i. This hindrance, in turn, promotes the cleavage of GB and the nucleation of intergranular nanovoids, as shown in Figs. \ref{Fig3}b-e, ultimately leading to intergranular fracture. This might result in a transition of transgranular-dominated fracture to intergranular-dominated fracture, as well as the hydrogen embrittlement. It agrees with the experimental result that the abundant H atoms at the GBs are crucial steps in causing the observed hydrogen-induced intergranular failure \cite{Wang2014a}.

In summary, we have developed and validated a new NNP in this work, building upon our previous Fe-H NNP. This new NNP, combined with GPU accelerators, demonstrates significantly improved efficiency, approximately 40 times faster than our previous n2p2 NNP, while maintaining similar accuracy to DFT calculations. To demonstrate the advantage of this NNP, we applied the NNP to investigate hydrogen embrittlement using through-thickness \{110\}$\langle 110 \rangle$ crack models and polycrystalline $\alpha$-iron models. It is revealed that the hydrogen atoms segregated at crack tip facilitate the brittle-cleavage failure of crack followed by crack growth, as well as those at GB promote the nucleation of intergranular nanovoids and thus intergranular fracture. We anticipate that this NNP will serve as a high-efficiency tool for gaining atomic-scale insights into hydrogen embrittlement.

\section*{CRediT authorship contribution statement}

S. H. Zhang: Methodology, Software, Visualization, Writing – original draft. F. S. Meng: Methodology, Writing – review \& editing. R. Fu: Methodology. S. Ogata: Conceptualization, Writing – review \& editing, Supervision, Funding acquisition.

\section*{Declaration of competing interest}

The authors declare that they have no known competing financial interests or personal relationships that could have appeared to influence the work reported in this paper.

%

\section*{Acknowledgement}

S.H.Z. and S.O. were funded by the JSPS Postdoctoral Fellowships for Research in Japan (Standard), the Grant-in-Aid for JSPS Research Fellow Grant No. 22F22056, and the JSPS KAKENHI Grant No. JP22KF0241. Part of the calculations were performed on the large-scale computer systems at the Cybermedia Center, Osaka University, and the Large-scale parallel computing server at the Center for Computational Materials Science, Institute for Materials Research, Tohoku University. S.O. acknowledges the support by the Ministry of Education, Culture, Sport, Science and Technology of Japan (Grant Nos. JPMXP1122684766, JPMXP1020230325, and JPMXP1020230327), and the support by JSPS KAKENHI (Grant No. JP23H00161).


\bibliography{Refs/Refs}

\newpage
\section*{Figures}

\newpage

\newpage

\newpage
\section*{Tables}
\input{Tables/table1}

\end{document}

%% file: Tables/Table1.tex
\begin{table}[H]
	
	\renewcommand\arraystretch{1.0}
	
	\centering
	\setlength{\tabcolsep}{10pt}
	
	\caption{Properties of Fe-H system produced by deepMD NNP, comparing to n2p2 NNP, DFT, and experimental results. $E_{i\text{-}110}^f$ is self-interstitial energy of the $\langle 110 \rangle$ dumbbell, which is the most stable self-interstitial atom configuration in $\alpha$-Fe \cite{NguyenManh2006}. $\langle 110 \rangle$ $\Sigma$3 and $\langle 001 \rangle$ $\Sigma$5 are the symmetric tilt GB with $\langle 110 \rangle$ and $\langle 001 \rangle$ tilt axes, respectively. The trapping energy of 6th H atoms at vacancy was calculated via the configuration of 2 H atoms at opposing O-sites while the other 4 H atoms at T-sites in one plane \cite{Meng2021}. Adsorption energy of 1 H atom on surface is calculated via ${E_{\rm{slab + H}}} - {E_{\rm{slab}}} - {E_{{\rm{H_2}}}}/2$, where $E_{\rm{slab + H}}$ and $E_{\rm{slab}}$ indicate the energy of the slab with and without H atoms on surface, respectively, and $E_{\rm{H_2}}$ is the energy of isolate H$_2$ molecule. $b$ in the energy of dislocation is the magnitude of Burgers vector $\bm{b}$ of dislocation. Kink nucleation energy of dislocation with 1 H atom at E$_k$ site (see Ref. \cite{Meng2021}) near screw dislocation was calculated.}
	
	\label{table1}
	
	\begin{threeparttable}
		
		\resizebox{0.6\width}{!}{
		\begin{tabular}{p{4.5cm}p{2.3cm}p{2.3cm}p{2.3cm}p{2.8cm}p{2.0cm}}
		\toprule
		Properties                                                                                  &                                     & deepMD NNP    & n2p2 NNP \cite{Meng2021} & DFT                                     & Exp                           \\
		\midrule
		Lattice constants (Å)                                                                       & $a_0$                                  & 2.834         & 2.83               & 2.83 \cite{Meng2021}                                              & 2.855 \cite{Basinski1955}               \\
		Elastic constants (GPa)                                                                  & $C_{11}$, $C_{12}$, $C_{44}$                       & 280, 128, 104 & 296, 147, 96       & 297, 151, 105 \cite{Meng2021}                                    & 239.5, 135.7, 120.7 \cite{Adams2006} \\
		Bonding energy of H$_2$ (eV)       &       &    4.50            &     4.54                    &   \\
		Surface energy (J/m$^2$)                                                                    & (001)                                  & 2.501         & 2.479              & 2.488 \cite{Meng2021}                                            &                               \\
		& (110)                                & 2.461         & 2.436              & 2.449 \cite{Meng2021}                                            &                               \\
		& (111)                                & 2.686         & 2.695              & 2.691 \cite{Meng2021}                                           &                               \\
		& (112)                                & 2.601         & 2.586              & 2.575 \cite{Meng2021}                                             &                               \\
		Vacancy formation energy (eV)                                                            & monovacancy            & 2.327         & 2.203              & 2.223 \cite{Meng2021}                                             & 2.0 \cite{DeSchepper1983}                 \\
		Self-interstitial energy (eV)                                                            & $E_{i\text{-}110}^f$ & 4.2           & 4.037              & 4.023 \cite{Meng2021}                                             &                               \\
		Unstable SFE (J/m$^2$)                                                                      & (112)$\langle111\rangle$   & 1.02          & 1.17               &                                                   &                               \\
		Tilt GB (J/m$^2$)                                                               & $\langle 110 \rangle$ $\Sigma$3  & 0.42          & 0.46               &                                                   &                               \\
		& $\langle 001 \rangle$ $\Sigma$5  & 1.35          & 1.51               &                                                   &                               \\
		Interstitial H atom (eV)                                                                   & T-site                              & 0.267         & 0.236              & 0.22 \cite{Geng2017}, 0.234 \cite{Kuopanportti2016}, 0.21 \cite{Counts2010}   &                               \\
		& O-site                              & 0.411         & 0.389              & 0.35  \cite{Chen2017}                                            &                               \\
		Diffusion barrier of H atom (eV) &      T to T-site           & 0.099         & 0.108              & 0.092 \cite{Meng2021}, 0.088 \cite{Jiang2004}, 0.096 \cite{He2017} &                               \\
		Trapping energies of H at vacancy (eV)                                                   & 1st H                                  & 0.713         & 0.592              & 0.584 \cite{Hayward2012}, 0.575 \cite{Ohsawa2012}, 0.616 \cite{Hayward2013} &                               \\
		& 6th H                                  & 0.035         & 0.046              &                                                   &                               \\
		Adsorption energy of 1 H atom on surface (eV) & (001)                                  & 0.49          & 0.452              & 0.46 \cite{Wang2014}                                    &                               \\
		Energy of dislocation relative to easy core (meV/$b$)                                       & Hard core                           & 51.5          & 47.4               & 39.3 \cite{Itakura2012}, 33.2 \cite{Dezerald2014}, 57.9 \cite{Wakeda2017}    &                               \\
		& Split core                          & 69.9          & 82.3               & 108 \cite{Itakura2012}, 87.9 \cite{Dezerald2014}, 110.3 \cite{Wakeda2017}    &                               \\
		& Saddle point                        & 34.8          & 38.2               & 37.9 \cite{Itakura2012}, 34.9 \cite{Dezerald2014}, 49.2 \cite{Wakeda2017}    &                               \\
		Kink nucleation energy (eV)                                                              & Without H                           & 0.7           & 0.7                &                                                  &                               \\
		& With H                              & 0.62          & 0.67               &                                                  &                             \\ 
		\bottomrule  

\end{tabular}
}

\end{threeparttable}

\end{table}